\documentclass[aps,prb,twocolumn,showpacs,superscriptaddress]{revtex4}

\usepackage{dcolumn, xspace,units}
\usepackage{fdsymbol}
\usepackage{graphicx}
\usepackage{bm}
\usepackage[utf8]{inputenc}
\usepackage{color}

\newcommand{\un}[1]{\ensuremath{\,\textrm{#1}}}
\newcommand{\vg}{\ensuremath{V_{\textrm{g}}}}
\newcommand{\vsd}{\ensuremath{V_{\textrm{sd}}}}
\newcommand{\didv}{\ensuremath{\textrm{d}I/\textrm{d}\vsd}}
\newcommand{\eabs}{\ensuremath{\varepsilon_\text{abs}}}
\newcommand{\Dprobe}{\ensuremath{\Delta_\text{probe}}}
\newcommand{\DNb}{\ensuremath{\Delta_\text{Nb}}}
\newcommand{\DAl}{\ensuremath{\Delta_\text{Al}}}

\begin{document}

\title{Temperature dependence of Andreev spectra in a superconducting carbon
nanotube quantum dot}

\author{A. Kumar}
\altaffiliation[Present address: ]{School of Materials Science and Technology, 
Indian Institute of Technology (BHU), Varanasi-221005, India}
\affiliation{Institute for Experimental and Applied Physics, University of
Regensburg, 93040 Regensburg, Germany}

\author{M. Gaim}
\affiliation{Institute for Experimental and Applied Physics, University of
Regensburg, 93040 Regensburg, Germany}

\author{D. Steininger}
\affiliation{Institute for Experimental and Applied Physics, University of
Regensburg, 93040 Regensburg, Germany}

\author{A. Levy Yeyati}
\affiliation{Departamento de F\'{\i}sica Te\'orica de la Materia Condensada,
Condensed Matter Physics Center (IFIMAC) and Instituto Nicol\'as  
Cabrera, Universidad Aut\'onoma de Madrid, E28049 Madrid,
Spain
}

\author{A. Mart\'{i}n-Rodero }
\affiliation{Departamento de F\'{\i}sica Te\'orica de la Materia Condensada,
Condensed Matter Physics Center (IFIMAC) and Instituto Nicol\'as  
Cabrera, Universidad Aut\'onoma de Madrid, E28049 Madrid,
Spain
}

\author{A. K. H\"{u}ttel}
\altaffiliation{E-mail: andreas.huettel@ur.de}
\affiliation{Institute for Experimental and Applied Physics, University of
Regensburg, 93040 Regensburg, Germany}

\author{C. Strunk}
\altaffiliation{E-mail: christoph.strunk@ur.de}
\affiliation{Institute for Experimental and Applied Physics, University of
Regensburg, 93040 Regensburg, Germany}
\date{\today}

\begin{abstract}
Tunneling spectroscopy of a Nb coupled carbon nanotube quantum dot reveals the
formation of pairs of Andreev bound states (ABS) within the superconducting
gap. A weak replica of the lower ABS is found, which is generated by
quasi-particle tunnelling from the ABS to the Al tunnel probe. An inversion of
the ABS-dispersion is observed at elevated temperatures, which signals the
thermal occupation of the upper ABS. Our experimental findings are well
supported by model calculations based on the superconducting Anderson model.
\end{abstract}


\maketitle

\section{Introduction}

The proximity effect in a superconductor coupled to a mesoscopic normal
conductor leads to a wide range of new quantum phenomena. These include Andreev
reflection at normal and superconductor interfaces, the formation of Andreev
bound states (ABS) in confined geometries, and proximity-induced supercurrent
flow through normal conductors.\cite{Likharev79, Doh05, Jarillo06, nanosquid,
Pallecchi08} The recent experimental detection of individual Andreev bound
states\cite{Pillet10,Dirks11} as well as the efforts towards demonstrating
Majorana states in superconductor coupled nanostructures devices with strong
spin-orbit interaction\cite{Mourik12} received significant experimental and
theoretical interest and opened a new area of research.

Quantum dots (QDs) coupled to superconducting leads provide an ideal system to
test the theoretical predictions on the interplay of ABS with the Kondo 
effect.\cite{Kim13} A 0 to $\pi$ junction transition\cite{Martin11} has been 
reported in S-QD-S systems by measuring the sign reversal (positive to 
negative) of the Josephson supercurrent for even to odd occupation of the 
QD.\cite{Franceschi10, Maurand12, Dam06} Theoretical calculations suggest that 
this quantum phase transition in the S-QD-S Josephson junction devices 
is signaled also by the crossing of two Andreev levels.\cite{Vecino03, Lim08}
Depending on the ratio of the Kondo temperature $T_K$ at the center of a 
Coulomb valley and the superconducting gap energy $\Delta$, the ABS display a 
crossing ($k_B T_K \ll \Delta$) or a non-crossing ($k_B T_K \gg \Delta$) 
dispersion $\eabs(\vg)$ as a function of gate voltage $\vg$. These predictions 
have been confirmed by recent experimental studies using Al-contacted 
semiconductor quantum dot / nanowire devices.\cite{Kim13, Deacon10, Chang12, 
Lee13}

It has also been predicted theoretically that there could be up to four ABS for 
a single level model in the superconducting gap. However the two outer ABS may 
not be visible in the transport spectrum since they can merge with the 
continuum.\cite{Vecino03, Martin12, Yoshioka00, Meng09} So far most of the
experimental studies of ABS formation are limited to Al as contacts material,
which restricts these experiments to very low temperatures and magnetic fields.
Despite these earlier reports, the study of hybrid nanostructures with larger
gap superconducting elements is required to allow for a more complete
understanding of ABS formation in these devices.

Here, we report on low temperature tunnelling spectroscopy measurements on an 
individual carbon nanotube quantum dot device strongly coupled to a Nb 
superconducting loop and weakly coupled to an Al tunnel probe. Two types of ABS 
are observed in Coulomb valleys with different charging energy. In some gate 
regimes two pairs of ABS are found within the superconducting gap. In addition, 
next to the main ABS conductance resonance a weaker conductance peak is 
present, which is interpreted as quasi-particle tunnelling from the ABS
to the Al tunnel probe. At higher temperatures tunneling from the thermally
populated upper ABS becomes visible and shows an opposite curvature at the
center of the Coulomb valleys. Calculations based on the superconducting
Anderson model are used to describe the experimentally observed subgap
features.

\section{Experimental details}

Single wall CNTs are grown on a Si/SiO$_2$ substrate by chemical vapor
deposition using a Fe/Mo based catalyst and methane as precursor gas. The highly
doped Si substrate with 300 nm SiO$_2$ layer serves as a global back gate. 
\begin{figure}[th]
\includegraphics[width=9cm]{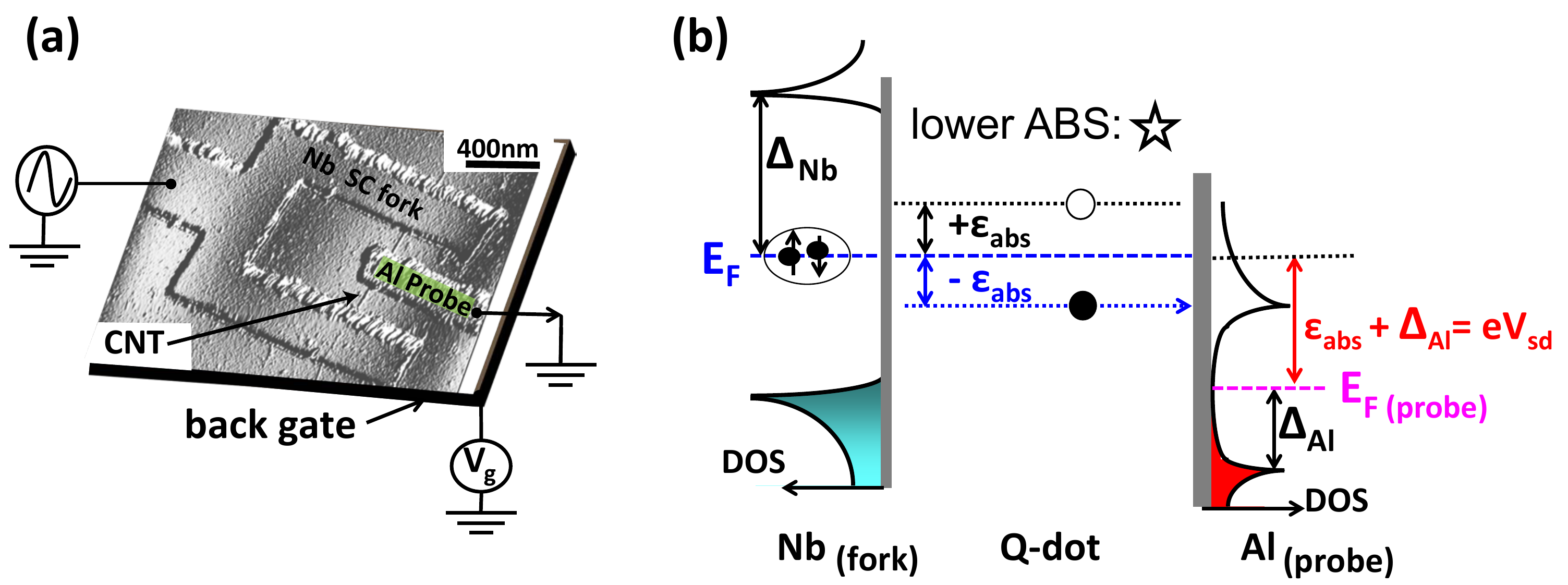}
\caption{(Color online)
(a) Atomic force micrograph of a typical device. The niobium fork is strongly,
the tunnel probe weakly coupled to the nanotube. (b) Manifestation of ABS at 
energies $e \vsd = \pm (\eabs + \DAl$): a  peak in the differential conductance 
is observed when unoccupied (occupied) states of the tunnel probe are aligned 
to the energy ($\mp\eabs$) of the lower (upper) state of an ABS pair.
}\label{fig:device}
\end{figure}
A niobium ($3 \un{nm}$ Pd / $60\un{nm}$ Nb) superconducting loop and a tunnel
probe ($1\un{nm}$ Ti / $60\un{nm}$ Al) are patterned using standard electron
beam lithography on an individual single wall carbon nanotube (Figure
\ref{fig:device}(a)). A $3\un{nm}$ Pd interlayer is used to improve the 
coupling between the superconducting fork and the nanotube and thereby increase 
the superconducting proximity effect. For weak coupling of the tunnel probe to 
the CNT a $1\un{nm}$ thin Ti adhesion layer is used. Low temperature electrical
transport measurements are performed in a $^3$He/$^4$He dilution refrigerator
with heavily filtered signal lines down to $\geq 28\un{mK}$. The
differential conductance $\didv$ of the superconducting tunnel probe weakly
connected to the nanotube was measured employing conventional lock-in
techniques by adding a low frequency ac excitation voltage ($V_{ac} = 
5\,\mu\text{V}$, $f\simeq 137\un{Hz}$) onto the dc bias voltage $\vsd$. 
Variation of the back gate voltage gives access to the ABS spectrum in the 
CNT-QD. 

The device parameters are extracted by measurements in both the
superconducting and the normal conducting state (by applying a magnetic field)
of the contacts.
\begin{figure}[th]
\includegraphics[width=8cm]{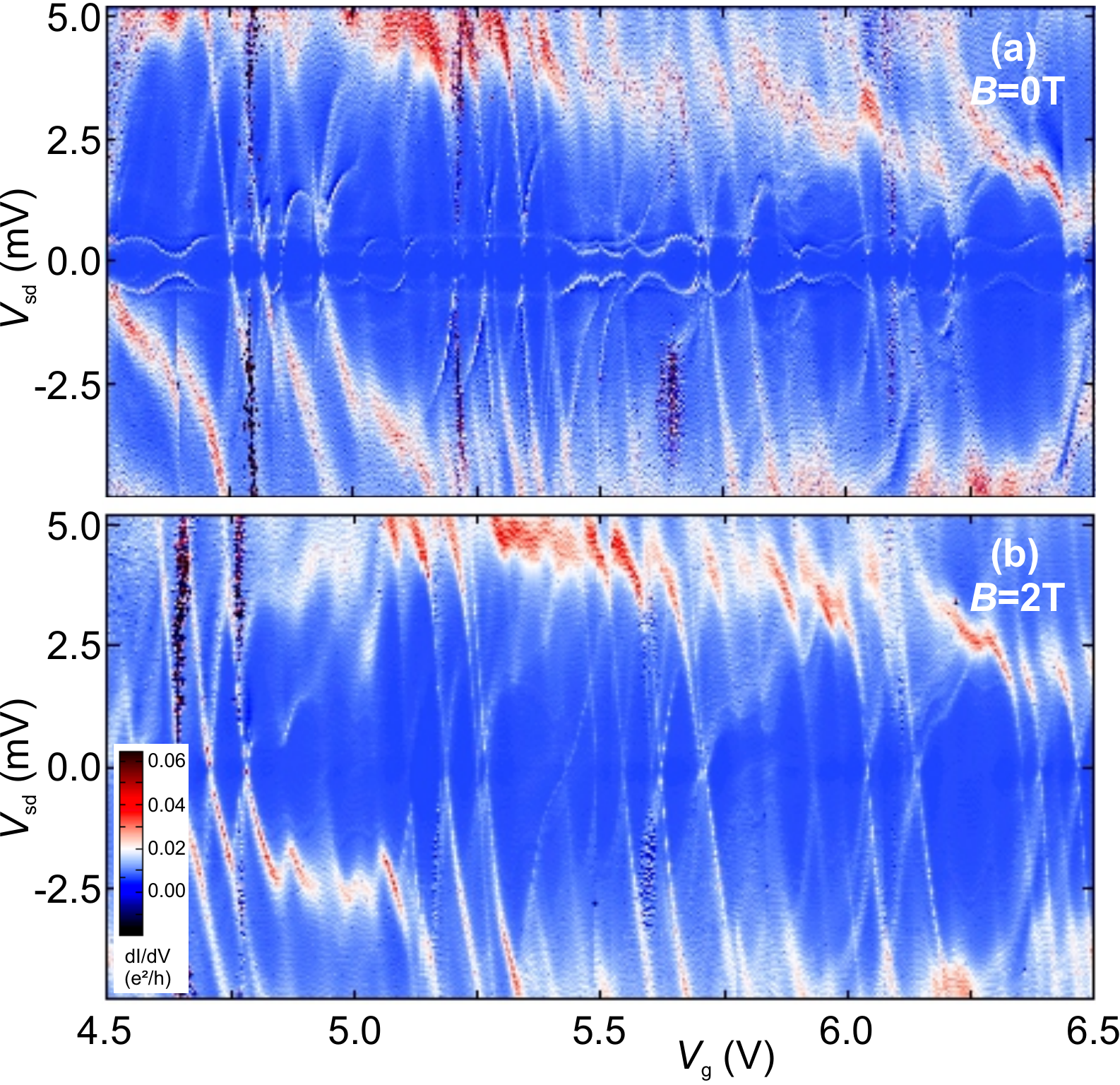}
\vspace*{-0.3cm}
\caption{
Overview measurements of the differential conductance $\didv(\vg,\vsd)$ as 
function of back gate voltage \vg\ and bias voltage \vsd, for (a) $B=0\un{T}$ 
and (b) $B=2\un{T}$. While the superconducting energy gap in the contacts is 
clearly visible in (a), in (b) superconductivity is suppressed and normal-state 
Coulomb blockade behaviour emerges.
}\label{fig:overview}
\end{figure}
Corresponding overview plots of the differential conductance are shown in 
Fig.~\ref{fig:overview}. While in Fig.~\ref{fig:overview}(a) at zero applied 
magnetic field features pertaining to superconductivity in the leads dominate 
transport at low bias, a magnetic field of $B=2\un{T}$ restores the typical 
pattern of Coulomb blockade in Fig.~\ref{fig:overview}(b). The 
superconducting gap of the Nb loop is found to be $\Delta_\text{Nb} \simeq 
0.450\un{meV}$, and for the Al tunnel probe $\Delta_\text{Al} \simeq 
0.165\un{meV}$ and $T_{c,\text{Al}} \simeq 1.12\un{K}$ are estimated. Typical 
parameter values of the CNT-QD are the charging energy $E_c \simeq 2.5-6 
\un{meV}$, the tunnel coupling to the leads $\Gamma \simeq 0.5-1\un{meV}$, and 
the coefficient $\alpha_g \simeq 0.054$ which relates the energies 
$\varepsilon=\alpha_ge\vg$ of the quantum dot states to $\vg$.

\section{Andreev Bound State features}

\begin{figure}[t]
\includegraphics[width=8cm]{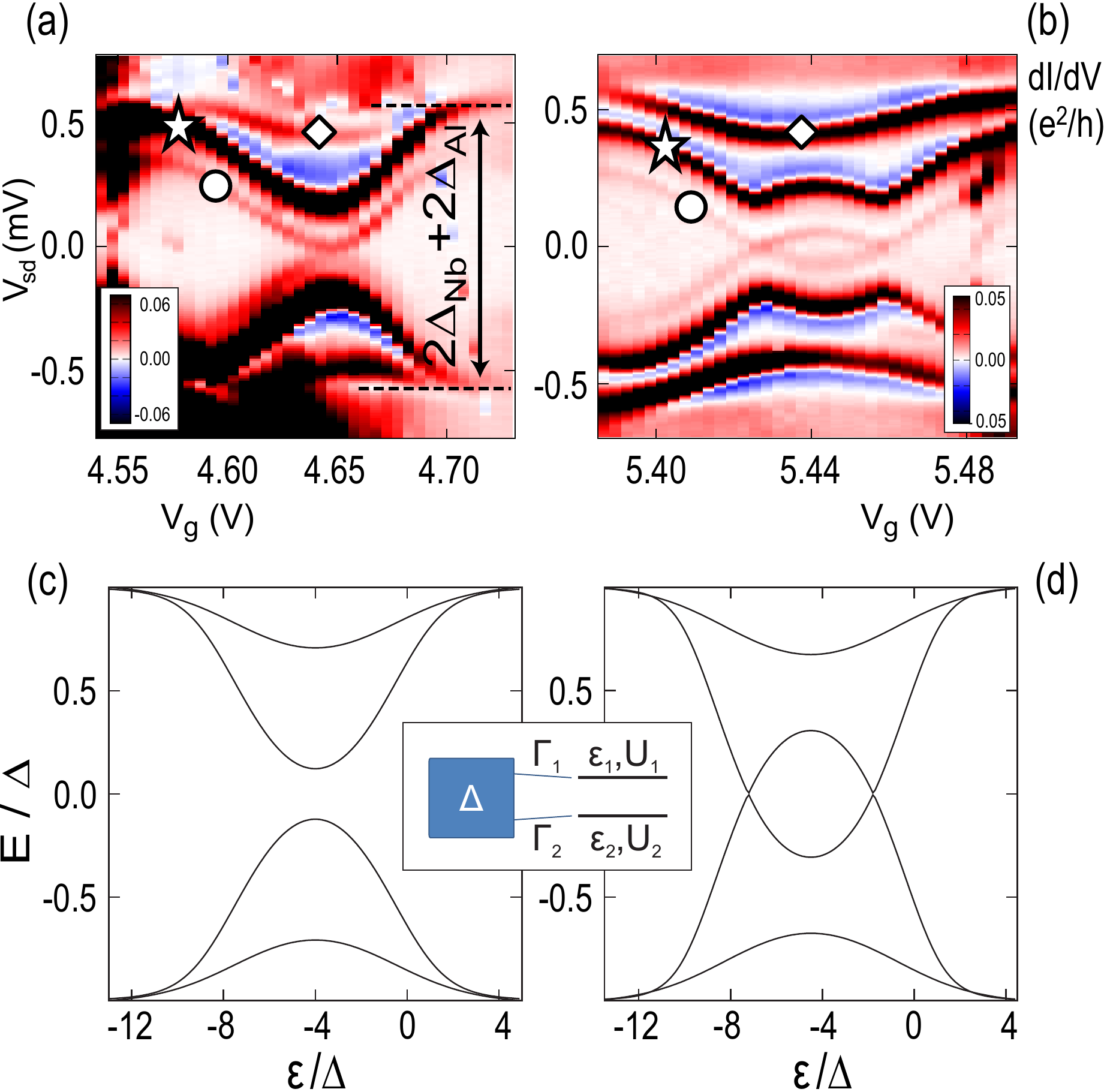}
\vspace*{-0.3cm}
\caption{(a), (b) Differential conductance plotted as a function of bias 
voltage $\vsd$ and back gate voltage $\vg$ for two different gate regimes, (a) 
$\vg \simeq 4.65\un{V}$ and (b) $\vg \simeq 5.44\un{V}$. The sharp resonances 
are the signature of the Andreev bound states: a main resonance of high 
conductance ($\medwhitestar$), a weak conductance resonance ($\medcircle$) 
running parallel to the main resonance peak ($\medwhitestar$), and a third 
additional resonance close to the gap edge ($\Diamond$). (c), (d) ABS spectrum 
calculated using NRG for a two channel superconducting Anderson model with 
parameters $E_c = 8\Delta,\Gamma_1 =1.85\Delta, \Gamma_2=4\Delta$ and $E_c = 
9\Delta, \Gamma_1 =1.2\Delta, \Gamma_2 =4\Delta$, respectively. Inset: 
schematic representation of the theoretical model used to describe the main 
multilevel features, see text. Two independent transport channels connect the 
quantum dot with the superconducting reservoir.
 }\label{fig:stability}
\end{figure}
\begin{figure*}[tbh]
\includegraphics[width=16cm, height=8cm]{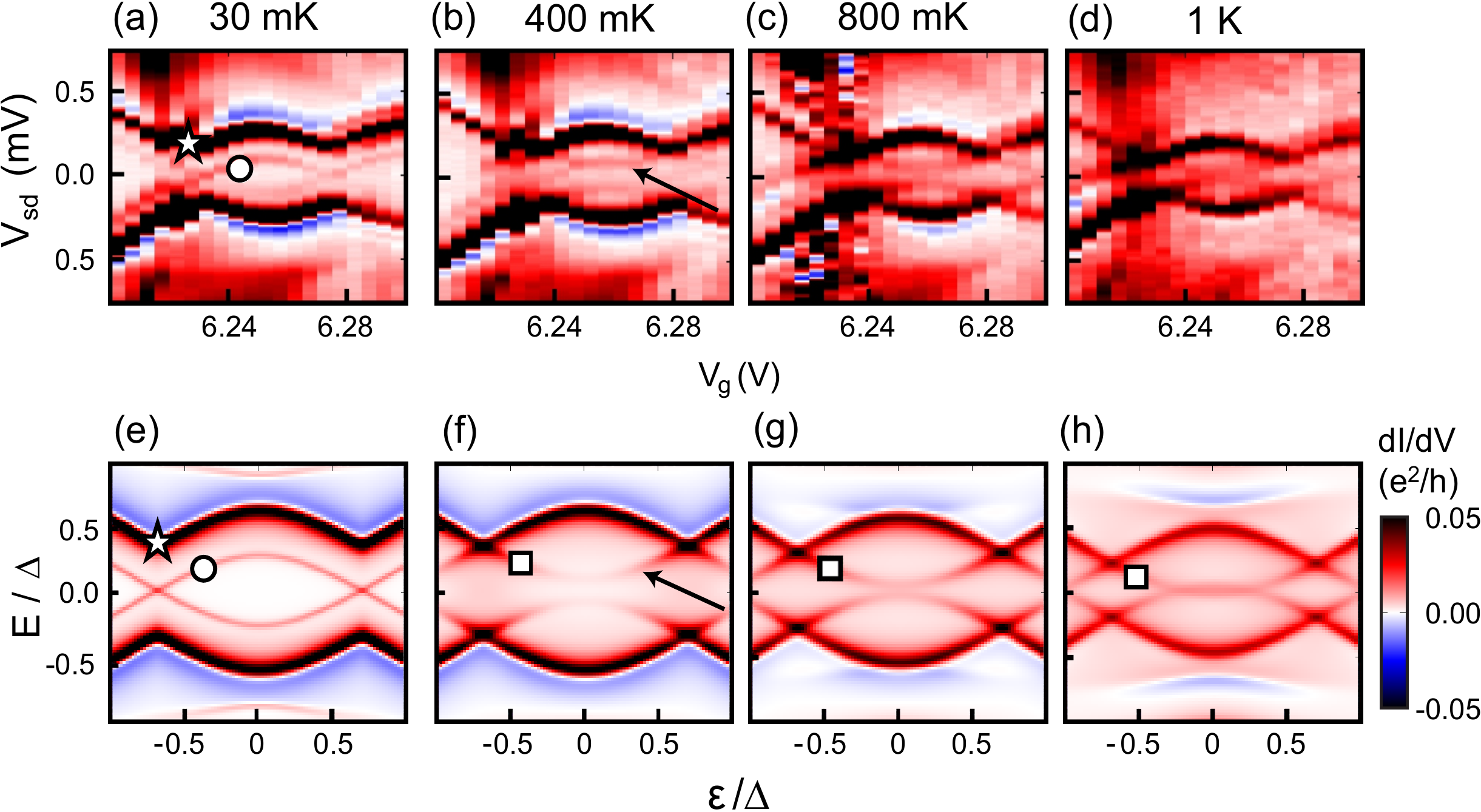}
\vspace*{-0.3cm}
\caption{(a)--(d) Two dimensional differential conductance maps of the tunnel
probe as function of tunnel probe bias $\vsd$ and back gate voltage $\vg$ taken
at indicated temperatures. The two subgap resonance lines are the main ABS
($\medwhitestar$) resonance (higher conductance) and a secondary weaker
conductance ($\medcircle,\square$) resonance running parallel to the main
resonance at $T\lesssim 200\,$mK. Figure (e)-(h) are model calculations 
that include the effect of populating the upper ABS according to the 
temperatures in (a)-(d).
}\label{fig:temperature}
\end{figure*}
Figure \ref{fig:stability} shows two details of the stability diagram, where 
the differential conductance \didv\ is again plotted as a function of bias 
\vsd\ and gate voltage \vg, at base temperature and zero magnetic field. The 
two panels (a) and (b) correspond to two different gate voltage ranges 
exhibiting different charging 
energies  $\simeq 4.9\un{meV}$ and $\simeq 5.7\un{meV}$, respectively. Within 
the superconducting gap range $\left|\vsd\right|<\DNb+\DAl$ there are three 
subgap features: two main resonances of high conductance ($\medwhitestar$, 
$\Diamond$), and a weak conductance resonance ($\medcircle$) running parallel 
to the resonance ($\medwhitestar$). Strong peaks in the differential 
conductance measurements are expected when an Andreev level at $\pm\eabs$ is 
aligned to the BCS singularity of the density of states  of the tunnel probe 
(see figure \ref{fig:device}(b)). This results in pronounced conductance peaks 
at voltages $\vsd=(\eabs+\DAl)/e$.\cite{Pillet10} The weak conductance peak 
($\medcircle$) running parallel to the lower ABS  ($\medwhitestar$) at lower 
bias voltages is understood as a replica of the ABS at low temperatures (see 
below).

As already mentioned the ABS ($\medwhitestar$) spectrum for odd charge states 
can show non-crossing (Fig. \ref{fig:stability}(a)) or crossing behavior of the 
pairs of bound states (Fig. \ref{fig:stability}(b)), resulting in a 0-$\pi$
quantum phase transition.\cite{Deacon10, Kim13} This is controlled by the 
ratio $T_K(E_c,\Gamma)/\Delta$.\cite{Kim13, Deacon10, Chang12, Pillet13, 
Futterers13} In case of non-crossing the system always stays in the 0-phase, 
i.e. a (singlet) ground state whereas for a zero bias crossing of the ABS the 
system changes its ground state from 0 to $\pi$-state (magnetic doublet).

Fig.~\ref{fig:stability}(c) and (d) show the ABS spectrum obtained from the
superconducting Anderson model using the numerical renormalization group (NRG)
method.\cite{Yoshioka00,Martin12} We have found that the presence of two pairs 
of ABS can be explained by assuming the presence of two nearly degenerate 
levels, e.g., resulting from the often lifted KK'-symmetry that are coupled to 
two independent channels in the leads, as schematically drawn in the inset of 
the figure. For the case of Fig.~\ref{fig:stability}(a) 
both channels are in the 0-phase and the ABS exhibit a non-crossing behavior 
as correspondingly shown in Fig.~\ref{fig:stability}(c). On the other hand, in 
the case of Fig.~\ref{fig:stability}(b) the inner ABS exhibit a loop indicating 
the transition to the $\pi$-phase, which is accounted for in the theoretical 
result of Fig.~\ref{fig:stability}(d) by a larger $E_c/\Gamma$ ratio for one of 
the channels. The $U/\Delta$ values were taken as 8 
(Fig.~\ref{fig:stability}(c)) and 9 (Fig.~\ref{fig:stability}(d)) close to the 
experimental estimations while the parameters $\Gamma_{1,2}$ were chosen to get 
a qualitative fit of the experimental results.

\section{Temperature dependence}
To understand the origin of the weak replicas of the inner pair of ABS
$(\medcircle)$ in our experiment, we investigate their dependence on 
temperature.  Figure~\ref{fig:temperature}(a--d) shows a detail of the 2D 
stability diagram in a different Coulomb valley at the indicated temperatures. 
In this gate regime, again a weak conductance resonance ($\medcircle$) running 
parallel to a pair of ABS ($\medwhitestar$) is observed. The apparent replica 
of the ABS can be understood if we assume that the probe DOS is finite at the 
Fermi energy $E_{F\text{(probe)}}$ and $E_{F\text{(probe)}}$ is aligned with 
the ABS: $e\vsd(\vg)=\pm\eabs(\vg)$. 
\begin{figure}[tbh]
\includegraphics[width=9cm]{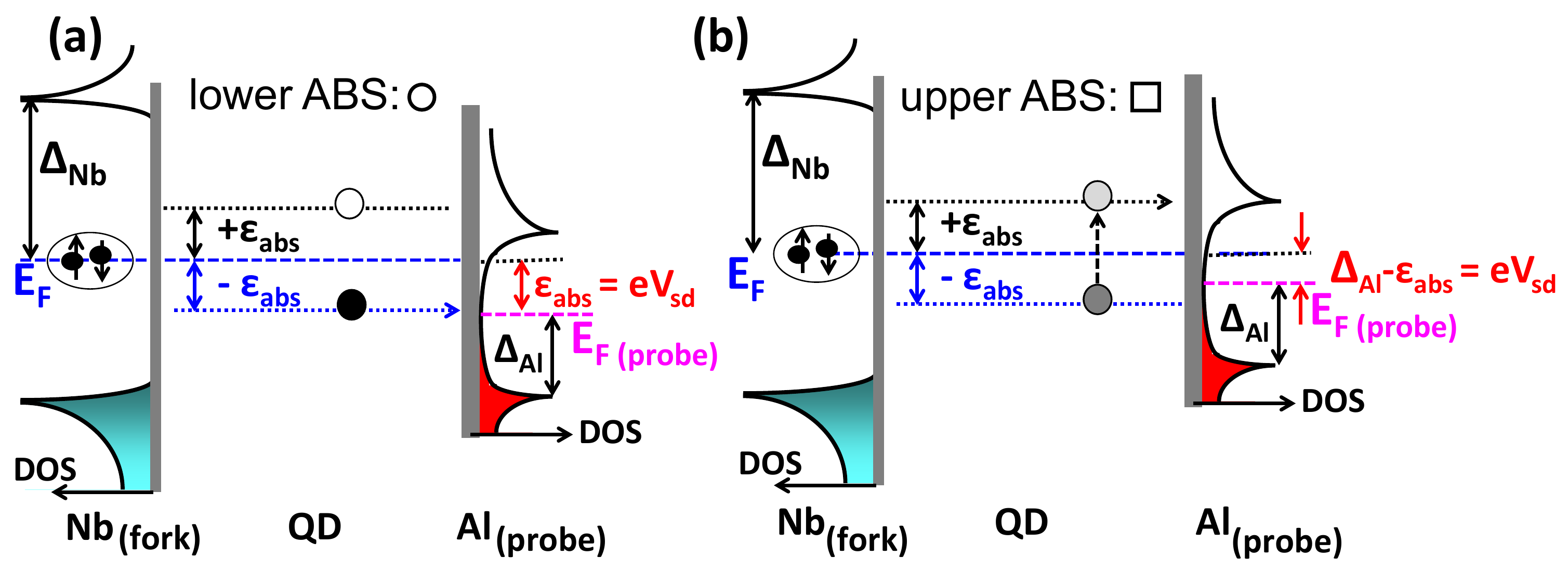}
\vspace*{-0.3cm}
\caption{(Color online) 
(a) A replica of the lower ABS ($\medcircle$) at $e \vsd = \pm \eabs$
is observed when the probe DOS is finite at the Fermi energy
$E_{F\text{(probe)}}$ and $E_{F\text{(probe)}}$ is aligned
with the ABS.  (b) At higher temperatures also the upper ABS ($\square$) at
$+\eabs$ is thermally populated.  In this case a peak at $e\vsd = \pm(\DAl -
\eabs$) is observed in the differential conductance, where an unoccupied
(occupied) DOS of the tunnel probe is aligned to a thermally populated upper
(lower) ABS.
}\label{fig:sketch}
\end{figure}
This level arrangement illustrated in Fig.~\ref{fig:sketch}(a). For a 
non-superconducting tunnel probe one would expect the disappearance of the 
replicas together with a shift in the main ABS due to the suppression of the 
probe gap $\DAl$. We have verified this by applying a small magnetic field ($B 
\simeq 60\un{mT}$) that drives the Al tunnel probe into the normal state: it 
suppresses the replica and a clear crossing of the main ABS ($\medwhitestar$) 
is then present (data not shown).

The main ABS peak ($\medwhitestar$) remains unchanged up to the maximum 
temperature $1\un{K}$ investigated in the present study. The satellite peaks 
($\medcircle$) also do not show any significant change up to $200\un{mK}$. At 
$400\un{mK}$, however, the dispersion of the gate voltage dependence of the 
satellite peaks flattens at the center of CB valley and an additional 
conductance resonance starts to emanate from the main ABS resonance (see the 
arrow in Fig.~\ref{fig:temperature}(b)). At even higher temperatures 
$>400\un{mK}$, the gate voltage dependence of the satellite peak changes its 
curvature at the center of the Coulomb valley compared to the low temperature 
($<400\un{mK}$) case. 

As illustrated in the schematic of Fig.~\ref{fig:sketch}(b), we expect a peak 
in 
differential conductance when the maximum of the BCS density of states of the 
tunnel probe is aligned with the thermally populated upper state. The position 
of the secondary peak ($\square$) should then be inverted, and vary as 
$\vsd(\vg)=\pm(\Dprobe - \eabs(\vg)$). At intermediate temperatures ($\simeq 
400\un{mK}$) we observe that quasi particle tunnelling via both the lower and 
the thermally populated upper ABS  (Fig.~\ref{fig:temperature}(b)) contributes 
to the current. The combined contributions lead to a nearly flat dispersion of 
the secondary peak. At higher temperatures (Fig.~\ref{fig:temperature}(c,d)) 
transport of thermally excited quasi-particles via the upper ABS dominates, 
which leads to the opposite curvature of the dispersion in the center of the 
Coulomb valley. The slight shift of both the ABS and their replica towards 
smaller energies results from the reduction of $\DAl(T)$ as the temperature is 
increased. 

According to our model, the dispersion of all three types of conductance peak 
originates from the same dispersion relation $\eabs(\vg)$.  Hence, it should be 
possible to collapse the dispersion of the ABS and its replicas at low and high 
temperature on top of each other by suitable inversions and shifts. 
\begin{figure}[thb]
\includegraphics[width=8cm]{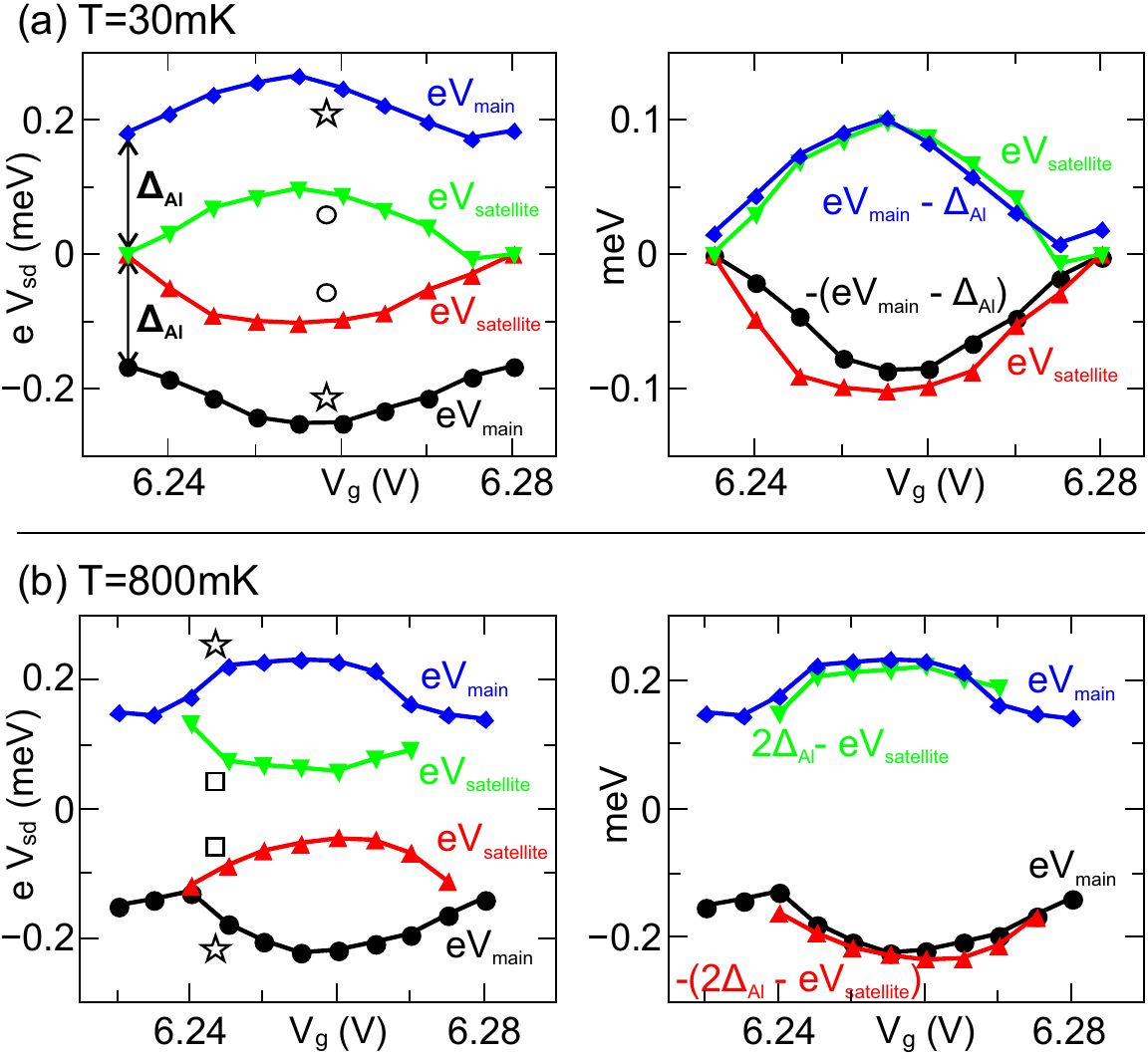}
\vspace*{-0.3cm}
\caption{(Color online)
Left: Bias positions of the main ABS and its satellite peaks, plotted as a 
function of gate voltage \vg\ for (a) $T=30\un{mK}$ and (b) $T=800\un{mK}$. 
Right: Plots combining the data of the left column graphs to demonstrate that 
the gate dependence of the peak positions can in each case be reduced to a 
single dispersion relation $\eabs(\vg)$; see the main text for details.
}
\label{fig:gatedep}
\end{figure}
This is illustrated in Fig.~\ref{fig:gatedep}, confirming our interpretation:
At low temperature $T=30\un{mK}$, Fig.~\ref{fig:gatedep}(a) left panel, we 
observe in the differential conductance measurement the main ABS at $e 
V_\text{main}= \pm (\eabs + \DAl)$ and a satellite peak at $e 
V_\text{satellite} =\pm \eabs$. After subtracting the SC gap of the tunnel 
probe ($\DAl = 0.165\pm 0.005\un{meV}$) at $30\un{mK}$, the peak position of 
the main ABS peaks show a good overlap with the satellite peak 
(Fig.~\ref{fig:gatedep}(a), right panel). At higher temperature $T=800\un{mK}$, 
as depicted in Fig.~\ref{fig:gatedep}(b) left panel, the dispersion of the 
satellite peak is inverted. Its position follows a relation $e 
V_\text{satellite} =\pm (\DAl - \eabs)$ with gate voltage in the Coulomb 
blockade valley. To verify this gate dependence, we compute and plot $\pm 
[2\DAl - e V_\text{satellite}(\vg)] = \pm [2\DAl -\{\DAl - \eabs(\vg)\}] = \pm 
[\DAl + \eabs(\vg)] = e V_\text{main}(\vg)$ for a reduced 
$\DAl(800\un{mK})=0.14\un{meV}$, and find a very good match with the observed 
main peak position $e V_\text{main}(\vg)$, see Fig.~\ref{fig:gatedep}(b) right 
panel. A similar analysis for other temperatures also shows an excellent 
agreement with the proposed mechanism.

\section{Superconducting Anderson model}

In addition to the measurement results, calculated conductance patterns for 
a quantum dot structure as described here are shown in 
Fig.~\ref{fig:temperature}(e)-(h). The conductance calculations were performed 
using a mean field description of the superconducting Anderson model coupled to 
two superconducting leads with two different gap parameters \DNb\ and \DAl.
As discussed in Ref.~\onlinecite{Vecino03} the magnetic phase can be 
represented within a mean field description by introducing an exchange 
parameter $E_\text{ex}$ which produces a splitting of the dot energy levels for 
different spin orientations. In these calculations the coupling to the Nb 
superconducting leads is included nonperturbatively (i.e.\ to all orders in 
perturbation theory) while the coupling to the Al probe is introduced to the 
lowest order in perturbation theory. This approximation is justified by the 
weak coupling strength to the probe electrode, manifested by the small 
conductance values experimentally observed (of the order of $0.1 e^2/h$). More 
precisely the conductance was calculated using
$$
G(V)=\frac{e \Gamma_p}{h} \frac{\partial}{\partial V} 
\int \,\text{d}E 
\left(
f(E-V)-f(E)
\right) 
\rho_p(E-V) \rho_d(E)
$$
where $\rho_{p,d}(E)$ are the local densities of states at the probe electrode 
and at the dot coupled to the Nb leads, respectively, $f(E)$ is the Fermi 
distribution function and $\Gamma_p$ is the tunneling rate from the dot to the 
probe electrode. While $\rho_d(E)$ is calculated using the mean field 
approximation for the Anderson model coupled to the Nb superconducting lead as 
described in Ref.~\onlinecite{Vecino03}, for $\rho_p(E)$ we use a standard 
BCS-like density of states in which we introduce a phenomenological 
Dynes-parameter\cite{Dynes} of $\simeq 0.1 \DAl$ broadening the BCS density of 
states for the Al probe.

As can be observed in Fig.~\ref{fig:temperature}, the model calculations give a 
good description of the 
evolution of the weak subgap features with temperature, and once more confirm 
our interpretation in terms of different transport mechanisms. It should be 
noted that the only change in the model parameters when going from 
Fig.~\ref{fig:temperature}(e) to Fig.~\ref{fig:temperature}(h) is due to the 
reduction of $\DAl(T)$ with temperature, while we assume 
$\DNb(T)=\text{const.}$ in this temperature range. The comparatively high value 
of the Dynes-parameter is at present not understood. This may be an effect of 
the electromagnetic environment,\cite{pekola2010} to which superconducting 
quantum dot devices are known to be extraordinarily sensitive.

\section{Multi-loop structures and negative differential conductance}

\begin{figure}[tbh]
\includegraphics[width=8cm]{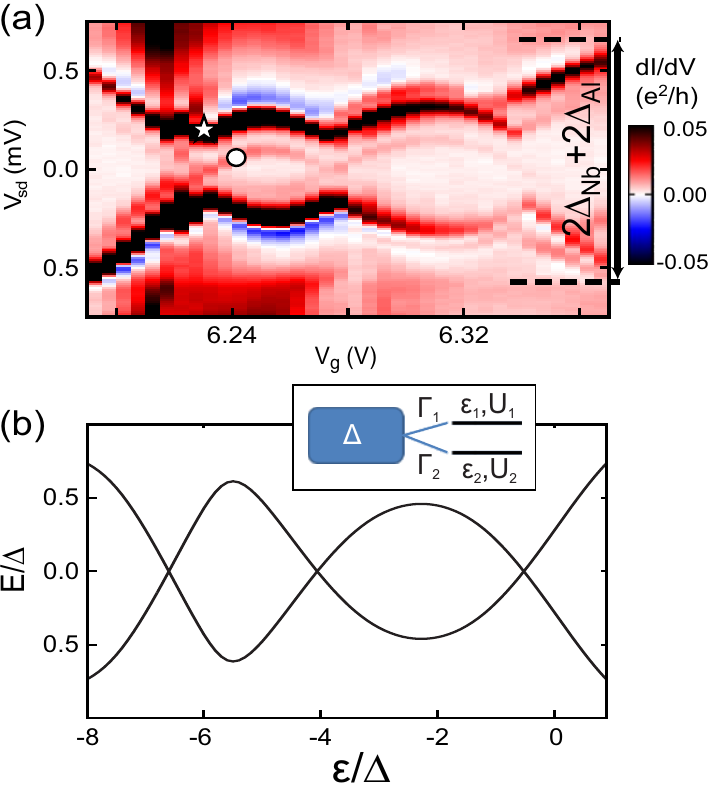}
\caption{
(a) Gate dependence of the differential conductance, extending 
Fig.~\ref{fig:temperature}(a) and displaying a multi-loop pattern. (b) ABS 
spectrum calculated by NRG assuming a single channel in the leads and using the 
parameters $E_c = 4.5\Delta, \Gamma_1 =0.5\Delta, \Gamma_2 =0.5\Delta, 
\varepsilon_1 - \varepsilon_1 =4.5\Delta$. Inset: schematic of the NRG model, 
see 
text. 
}\label{fig:multiloop}
\end{figure}
A peculiar feature of the present experimental results are the multiple loop 
ABS that are observed in a more extended gate range, as depicted in 
Fig.~\ref{fig:multiloop}. As opposed to the situation considered in
Fig.~\ref{fig:stability}(c) and (d), a theoretical description of such 
multi-loop features is obtained when more than one dot level couples to {\it the 
same} channel in the leads, see the inset of Fig.~\ref{fig:multiloop}(b). A NRG 
calculation considering such a situation with two hybridized levels on one dot 
can actually reproduce qualitatively the observed multi-loop patterns, as shown 
in  Fig.~\ref{fig:multiloop}(b). 

In several of the Coulomb valleys investigated in more detail the ABS-peaks are 
accompanied by a pronounced negative differential conductance (NDC), shown in 
blue color in Figs.~\ref{fig:stability}, \ref{fig:temperature}, and 
\ref{fig:multiloop}. NDC features have been predicted to appear due to the 
presence of the so-called Yu-Shiba-Rusinov (YSR) states for a QD with an odd 
number of electrons\cite{Andersen11} with highly asymmetric coupling to the 
leads. YSR-states can be regarded as a variant of ABS appearing for a magnetic 
impurity coupled to a superconductor,\cite{Shiba68} and their existence has 
been experimentally confirmed via scanning tunneling microscopy of magnetic 
atoms on superconducting surfaces\cite{Yazdani97} and QDs with superconducting 
leads.\cite{Kim13,Andersen11,Lee12,Yeyati97}

\section{Conclusions}

In conclusion, our transport spectroscopy of an individual carbon nanotube 
strongly coupled to wide gap Nb leads reveals several different types of
Andreev bound state spectra. Weak satellite peaks appear within the smaller 
probe superconducting gap which are a result of quasiparticle tunneling into a
residual density of states within this gap. At higher temperature these
satellite peaks change their dispersion as a function of gate voltage due to 
the thermal population of the upper state of an ABS pair. Our findings are well 
reproduced within the superconducting Anderson model in terms of combined NRG 
and mean field calculations. More efforts in this direction could be helpful 
also to discriminate Majorana bound states in similar hybrid nanostructures 
from other states at zero energy.

We gratefully acknowledge financial support from the Deutsche
Forschungsgemeinschaft within GRK 1570, SFB 689, the E.~Noether program
(Hu 1808-1), from Spanish MINECO through grant FIS2011-26516, and from the EU 
FP7 Project SE2ND. A.~K. thanks the Alexander von Humboldt Foundation for 
providing financial support during this research. We thank Kicheon Kang for 
enlightening discussions.

\end{document}